\documentclass[preprint]{emulateapj}
\usepackage{psfig}
\usepackage{graphicx}
\usepackage{apjfonts}

\newcommand{\sol}{$_{\odot}$}

\def\arcs{\hbox{$^{\prime\prime}$}}

\shorttitle{{\it Herschel} SPIRE Observations of SMM J163554.2+661225}
\shortauthors{FINKELSTEIN, K.D. ET AL.}

\begin{document}
\title{Probing the Star Formation History and Initial Mass Function of the z$\sim$2.5 Lensed Galaxy SMM J163554.2+661225 with {\it Herschel}\altaffilmark{*}}

\author{Keely D. Finkelstein\altaffilmark{1}\altaffilmark{,2}, Casey Papovich\altaffilmark{1}, Steven L. Finkelstein\altaffilmark{1}\altaffilmark{,2}\altaffilmark{,7}, Christopher N.A. Willmer\altaffilmark{3}, Jane R. Rigby\altaffilmark{4}, Gregory Rudnick\altaffilmark{5}, Eiichi Egami\altaffilmark{3}, Marcia Rieke\altaffilmark{3}, and J.-D. T. Smith\altaffilmark{6}}

\begin{abstract}
We present the analysis of {\it Herschel} SPIRE far-infrared (FIR) observations of the z = 2.515 lensed galaxy SMM J163554.2+661225.  Combining new 250, 350, and 500 $\mu$m observations with existing data, we make an improved fit to the FIR spectral energy distribution (SED) of this galaxy. We find a total infrared (IR) luminosity of L(8--1000 $\mu$m) = 6.9 $\pm$ 0.6$\times$10$^{11}$ L\sol; a factor of 3 more precise over previous L$_{IR}$ estimates for this galaxy, and one of the most accurate measurements for any galaxy at these redshifts.  This FIR luminosity implies an unlensed star formation rate (SFR) for this galaxy of 119 $\pm$ 10 M\sol~yr$^{-1}$, which is a factor of 1.9 $\pm$ 0.35 lower than the SFR derived from the nebular Pa$\alpha$ emission line (a 2.5$\sigma$ discrepancy). Both SFR indicators assume identical Salpeter initial mass functions (IMF) with slope $\Gamma=2.35$ over a mass range of 0.1 -- 100 M\sol, thus this discrepancy suggests that more ionizing photons may be necessary to account for the higher Pa$\alpha$-derived SFR. We examine a number of scenarios and find that the observations can be explained with a varying star formation history (SFH) due to an increasing star formation rate (SFR), paired with a slight flattening of the IMF.  If the SFR is constant in time, then larger changes need to be made to the IMF by either increasing the upper-mass cutoff to $\sim$200 M\sol, or a flattening of the IMF slope to 1.9 $\pm$ 0.15, or a combination of the two.  These scenarios result in up to double the number of stars with masses above 20 M\sol, which produce the requisite increase in ionizing photons over a Salpeter IMF with a constant SFH.

\end{abstract}


\altaffiltext{*}{{\it Herschel} is an ESA space observatory with science instruments provided by European-led Principal Investigator consortia and with important participation from NASA.}
\altaffiltext{1}{George P. and Cynthia Woods Mitchell Institute for Fundamental Physics and Astronomy, Department of Physics and Astronomy, Texas A\&M University, College Station, TX 77843-4242, USA} 
\altaffiltext{2}{Department of Astronomy and McDonald Observatory, University of Texas at Austin, 1 University Station C1400, Austin, TX 78712, USA}
\altaffiltext{3}{Steward Observatory, University of Arizona, 933 N. Cherry Ave., Tucson, AZ 85721, USA}
\altaffiltext{4}{NASA Goddard Space Flight Center, Code 665, Greenbelt, MD 20771, USA}
\altaffiltext{5}{Department of Physics and Astronomy, 1251 Wescoe Hall Dr., University of Kansas, Lawrence, KS 66045-7582, USA}
\altaffiltext{6}{Ritter Observatory, Department of Physics and Astronomy, University of Toledo, MS 113, Toledo, OH 43606, USA}
\altaffiltext{7}{Hubble Fellow}

\section{Introduction}
The recent launch of the {\it Herschel Space Observatory} (Pilbratt et al.\ 2010) has opened the high-redshift universe to detailed far-infrared (FIR) investigations. {\it Herschel} observations of galaxies are specifically useful as they constrain the peak of the FIR spectral energy distributions (SEDs) for z$\sim$2 galaxies, allowing robust measurements of the total infrared (IR) luminosity, L$_{IR}$ = L(8--1000 $\mu$m), and dust temperature (e.g., Amblard et al.\ 2010; Elbaz et al.\ 2010; Magdis et al.\ 2010).  From the total IR luminosity one can infer the star formation rate (SFR) and compare to SFRs based on other indicators. Both L$_{IR}$ and the dust temperature allow for detailed physical investigations of star-forming galaxies not possible without FIR observations.  

One important factor for understanding star-forming galaxies at high-redshift is the form of the initial mass function (IMF), specifically at the high-mass end. There is currently an ongoing debate as to whether the IMF is universal.  In nearby galaxies the IMF is well fit by a Salpeter (1955) IMF, where dN/dM $\propto$ M$^{-\Gamma}$, with slope $\Gamma$ = 2.35 (Elmegreen 2006; and references therein), or even steeper ($\Gamma >$ 2.35) in a few examples, such as in regions of the Large Magellanic Cloud (Gouliermis et al.\ 2005) and low surface brightness galaxies (Lee et al.\ 2004).  On the other hand, the summed IMF in clusters of galaxies appears to be Salpeter or slightly flatter ($\Gamma <$ 2.35) than Salpeter (e.g., Renzini et al.\ 1993; Lowenstein \& Mushotzky 1996).  Similarly, Baldry \& Glazebrook (2003) find that a slightly flatter IMF slope of $\Gamma$ = 2.15 is favored to fit the star formation history (SFH) and luminosity density in low redshift star-forming galaxies. Other evidence for a non-Salpeter IMF is shown by varying the IMF slope or extending the IMF upper-mass limit up to at least 120 M\sol~over the typical values of 100 M\sol~in order to reproduce the observed H$\alpha$ equivalent widths in Sloan Digital Sky Survey (SDSS) galaxies (Hoversten \& Glazebrook 2008; 2010).  Additionally, large Ly$\alpha$ equivalent widths in some Ly$\alpha$ emitting galaxies have been observed which cannot be explained by normal stellar populations, but may be caused by a top heavy (i.e., flatter) IMF slope (e.g., Kudritzki et al.\ 2000; Malhotra \& Rhoads 2002; Finkelstein et al.\ 2007). 

\begin{deluxetable*}{ccccc}
\tablecaption{SPIRE Flux Density Measurements for Component B$^{a}$}
\tablewidth{0pt}
\tablehead{
\colhead{Wavelength} & \colhead{Flux Density} & \colhead{Measured
  Error} & \colhead{Confusion Error$^{b}$} & \colhead{Total Quadrature-Summed Error$^{c}$}\\
\colhead{[$\mu$m]} & \colhead{[mJy]} & \colhead{[mJy]} &
\colhead{[mJy]} & \colhead{[mJy]}\\
}
\startdata
250&55.5&2& 5.8&6.1\\
350&56&7&6.5&9.4 \\
500&34&8&7.3&10.5
\enddata
\tablecomments{
\noindent$^{a}$\footnotesize Photometry for Component A is
approximately 1.5 times fainter.  $^{b}$\footnotesize Nguyen et al.\
(2010).  $^{c}$\footnotesize These total errors do not include an additional
15$\%$ error based on SPIRE's estimated absolute calibration uncertainty (Griffin et al.\ 2010).}
\end{deluxetable*}

One way to study the IMF is to compare SFR indicators based on nebular lines, whose luminosities are proportional to the number of ionizing photons generated primarily by O-stars, to those based on L$_{IR}$. In a dusty starburst most of the ionizing radiation is absorbed by dust and reprocessed into the thermal infrared, characterized by L$_{IR}$, where L$_{IR}$ is expected to be proportional to the bolometric luminosity (L$_{bol}$; Kennicutt 1998).  While {\it Herschel} can permit direct measurements of L$_{IR}$, the majority of galaxies at cosmological distances fall well below the 5$\sigma$ confusion limit of 25--35 mJy for {\it Herschel} at 250--500 $\mu$m (e.g.\ Chapman et al.\ 2002; Kneib et al.\ 2004).  Fortunately, gravitationally lensed systems can easily be observed in the sub-mm by {\it Herschel} and other sub-mm telescopes.  One such galaxy is SMM J163554.2+661225 at z = 2.515, which is gravitationally lensed by the rich cluster Abell 2218 (Kneib et al.\ 2004).  The system was first discovered in the sub-mm at 450 and 850 $\mu$m by SCUBA on the James Clerk Maxwell Telescope (JCMT; Kneib et al.\ 2004) as a multiply lensed system. The brightest component (component B) has a lensing magnification of 22 and the second brightest component (component A) has a lensing magnification of 14 (Kneib et al.\ 2004).  SMM J163554.2+661225 was observed by Rigby et al.\ (2008) in the mid-IR with {\it Spitzer} from 3.6 -- 8.0 $\mu$m with the Infrared Array Camera (IRAC; Fazio et al.\ 2004), at 24 -- 160 $\mu$m with the Multiband Imaging Photometry for {\it Spitzer},  (MIPS; Rieke et al.\ 2004), and with the {\it Spitzer} Infrared Spectrograph (IRS; Houck et al\ 2004).  Papovich et al.\ (2009, hereafter P09) also analyzed mid-IR {\it Spitzer} IRS spectroscopy of component B to make the first detection of Pa$\alpha$ emission in a high-redshift galaxy, from which they derived a revised dust--corrected Pa$\alpha$-SFR of 225 $\pm$ 37 M\sol~yr$^{-1}$ for a Salpeter IMF with an upper and lower mass cutoff of 100 M\sol~and 0.1 M\sol.

In this paper, we present new {\it Herschel} Spectral and Photometric Imaging Receiver (SPIRE; Griffin et al.\ 2010) observations of SMM J163554.2+661225. In \S 2 we analyze the SPIRE data, including the measurement of crowding-corrected flux densities. In \S 3 we derive the rest-frame FIR luminosity, dust temperature, and IR SFR.  We compare these results to local samples and other high-redshift sub-mm galaxies. In \S 4 we discuss the implications of the variations in SFRs on the star formation history and IMF.  In \S5 we present our summary and conclusions.  Where applicable, we use a cosmology with H$_{\mathrm{0}}$ = 70 km s$^{-1}$ Mpc$^{-1}$, $\Omega_{m}$ = 0.3 and $\Omega_{\lambda}$ = 0.7.

\section{Data Analysis}
{\it Herschel} SPIRE 250, 350, \& 500 $\mu$m data of Abell 2218 were taken as part of the {\it Herschel} Multi-tiered Extragalactic Survey (HerMES; Oliver et al.\ 2010).  The reduced SPIRE data of Abell 2218 were accessed through the HeDaM database\footnotemark, and were processed using the {\it Herschel} Interactive Pipeline Environment (HIPE  v2.3) as described in Oliver et al.\ (2010). The SPIRE 250, 350, and 500 $\mu$m images are blended for the two components (A and B, see Figure 1) of the lensed galaxy SMM J163554.2+661225. We used the galaxy fitting software GALFIT (v3.0 Peng et al.\ 2010) to perform point source function (PSF) fitting of the SPIRE data for the two components of the galaxy.   Figure 1 shows a 215\arcs~$\times$ 215\arcs~region around SMM J163554.2+661225 in {\it Spitzer} MIPS 24 $\mu$m, and the SPIRE 250, 350, and 500 $\mu$m data, with 250 $\mu$m contours overlaid on the other images.  We used the positions of the two components from the MIPS 24 $\mu$m image as a prior for the location of the components in the SPIRE 250 $\mu$m image.  We ran Source Extractor (Bertin \& Arnouts 1996) on the SPIRE images prior to GALFIT to estimate input magnitudes, radial profile, and positions of other sources in the image.  GALFIT was run on each of the SPIRE images on a 215\arcs$\times$215\arcs~region centered on the two dominant lensed components, fitting all sources.  GALFIT requires both an uncertainty image and a PSF.  The uncertainty images from the {\it Herschel} archive for this field were used, and we adopted the {\it Herschel}/SPIRE theoretical PSFs\footnotemark. The SPIRE 250, 350, and 500 $\mu$m beams are Gaussian with full width half maximum (FWHM) values of 18.1\arcs, 25.2\arcs, and 36.6\arcs, respectively. We also compared the results using model PSFs from the GOODS/{\it Herschel} survey (PI D. Elbaz). The flux density measurements varied by less than 5$\%$ based on the choice of PSF. 

\footnotetext[1]{http://hedam.oamp.fr}
\footnotetext[2]{http://herschel.esac.esa.int/twiki/bin/view/Public/SpireCalibrationWeb}

The components of the lensed galaxy are reasonably resolved in the 250 $\mu$m image (see figure~1).  
Therefore, we measured the  flux density for each source in the 250 $\mu$m image using GALFIT, and allowing the position to vary within the 1$\sigma$ positional uncertainty from SExtractor with no other constraints.  We measure a flux density at 250 $\mu$m for Component B (the brightest component), S$_{250}$ = 55.5 $\pm$ 2 mJy.  The measured flux density ratio between the two components is  S$_{250}$(B)/S$_{250}$(A) = 1.5 $\pm$ 0.1. This ratio is consistent with the flux density ratio measured at 24~\micron\ from P09, S$_{24}$(B)/S$_{24}$(A)=1.6, where the 24 $\mu$m flux densities of components A and B are 0.72 mJy and 1.16 mJy, respectively. This flux density ratio is also consistent with the reported ratios of components B and A at 450 and 850 $\mu$m (Kneib et al.\ 2004).  

The A and B components of this lensed galaxy are more severely blended
in the longer wavelength SPIRE 350 and 500 $\mu$m data (see figure~1).
Therefore, we add \textit{a priori} constraints on both the
astrometric positions and flux-density ratio to extract the individual
flux densities.  We measured flux densities for the components by
fixing the positions of each component to have the value measured in
the 250~\micron\ image.  We also include a constraint that the flux
density ratio at 350 and 500 $\mu$m for the two components have the
same ratio measured at 250 $\mu$m.  Given the uniformity in the
flux-density ratios of the A and B components from the mid--IR (24
$\mu$m), far-IR (250 $\mu$m) and sub-mm (450-850 $\mu$m) (see
paragraph above), this assumption seems valid.  Including these constraints with GALFIT, we derived flux densities at
350 and 500 $\mu$m for component B of S$_{350}$ = 56 $\pm$ 7 mJy, and
S$_{500}$ = 34 $\pm$ 8 mJy.  The values for component A are a
factor of 1.5  fainter (by construction).  The SPIRE flux density
measurements for component B are detailed in Table 1. 

In addition to the measured flux density uncertainties, SPIRE data
suffer noise from confusion, for which Nguyen et al.\ (2010) report
1$\sigma$ errors of 5.8, 6.5, and 7.3 mJy per beam for 250, 350, and
500 $\mu$m.  Therefore, we adopt conservative errors, adding the flux
measurement and confusion errors in quadrature, giving total flux
density errors of 6.1, 9.4, and 10.5 mJy. We use these total errors
when fitting the suite of models and computing IR luminosities and
dust temperatures as detailed below.  We also use previously published
MIPS 70, SCUBA 450 and 850 $\mu$m flux densities, which for Component
B are 2.56 $\pm$ 0.9 mJy, 75 $\pm$ 15 mJy, and 17 $\pm$ 2 mJy,
respectively (P09; Kneib et al.\ 2004).

\begin{figure}[!t]
\epsscale{1.15}
\plotone{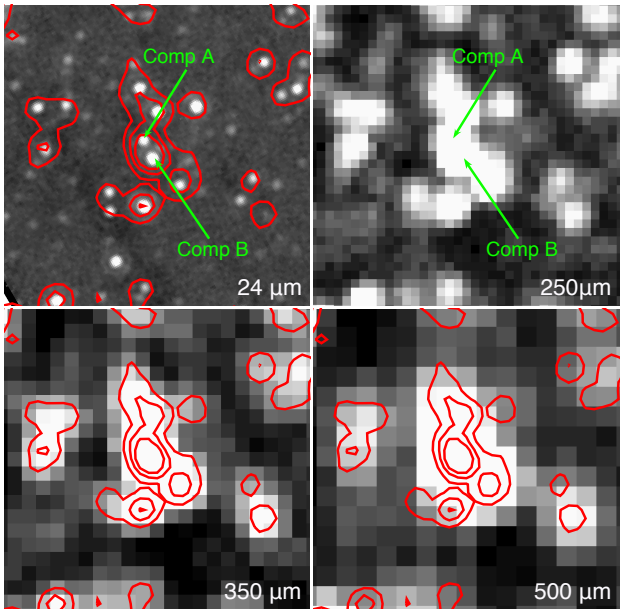}
\caption{Mid and Far-IR images of the lensed galaxy SMM J163554.2+661225.  The four panels show  $215\arcsec \times 215\arcsec$ regions around the two dominant lensed components (A and B, as labeled) in the {\it Spitzer} MIPS 24, {\it Herschel} SPIRE 250, 350, and 500 $\mu$m data (as labeled).   The contours show levels 1, 3, and 5 $\sigma$ above the noise per pixel in the 250 $\mu$m data.  While the 24$\mu$m data resolve components A and B of SMM J163554.1+661225, these components are blended in the \textit{Herschel} SPIRE data.   Accordingly, we fit the {\it Herschel} data with GALFIT using the positions of components A and B from the 24$\mu$m image.}
\vspace{3mm}
\end{figure}  

\section{Results}

\subsection{Infrared Luminosity and Dust Properties}

We use the {\it Herschel} SPIRE flux densities, {\it Spitzer} MIPS 70 $\mu$m data, and SCUBA 450 and 850 $\mu$m data to construct an infrared SED for Component B of SMM J163554.2+661225.  We fit a suite of IR SED templates from Chary \& Elbaz (2001; hereafter CE01) and Dale \& Helou (2002; hereafter DH02) to the combined data, finding the best fit model via $\chi^{2}$ minimization.  The total L$_{IR}$ is derived by integrating the best-fit models between 8 to 1000 $\mu$m.  From the best-fit DH02 template we derive a total infrared luminosity of 6.8$\times$10$^{11}$ L\sol, with a reduced $\chi$$^{2}$ = 1.4.  For the best-fit CE01 template we derive a total IR luminosity of L$_{IR}$ = 7.0$\times$10$^{11}$ L\sol, with a reduced $\chi$$^{2}$ = 1.2.  Both values of L$_{IR}$ are corrected for the gravitational-lensing magnification factor of 22.  Figure 2 shows the IR SED and the best-fit SED curves to the data. Figure 2 also shows the best-fit to the 70, 450, and 850 $\mu$m data alone.  Without the {\it Herschel} data the best fit template has a peak that is shifted to longer wavelengths.  To determine the uncertainty on our best-fit models we ran 10$^{4}$ Monte Carlo simulations, varying the input flux density measurements by a Gaussian random amount proportional to their errors in each simulation. The resulting 68$\%$ confidence ranges of the derived total IR luminosity from the best-fits are: 6.8 $\pm$ 0.6 $\times$10$^{11}$ L\sol~and 7.0 $\pm$ 0.5 $\times$10$^{11}$ L\sol.  Combining the results from the different templates we get an average L$_{IR}$ = 6.9 $\pm$ 0.6 $\times$10$^{11}$ L\sol; thus the uncertainty on L$_{IR}$ from the photometric data uncertainties dominate over systematic effects from the choice of models. Previous estimates of L$_{IR}$ ranged from 5 -- 10 $\times$10$^{11}$ L\sol~based on DH02, CE01, and Rieke et al.\ 2009 templates (P09) and 5.7 -- 9.5$\times$10$^{11}$ L\sol~based on DH02 templates (Rigby et al.\ 2008). Thus the addition of the {\it Herschel} data results in higher accuracy, and shrinks the uncertainty on L$_{IR}$ from a factor of order 2 down to 10$\%$.

\begin{figure}[t]
\epsscale{1.25}
\plotone{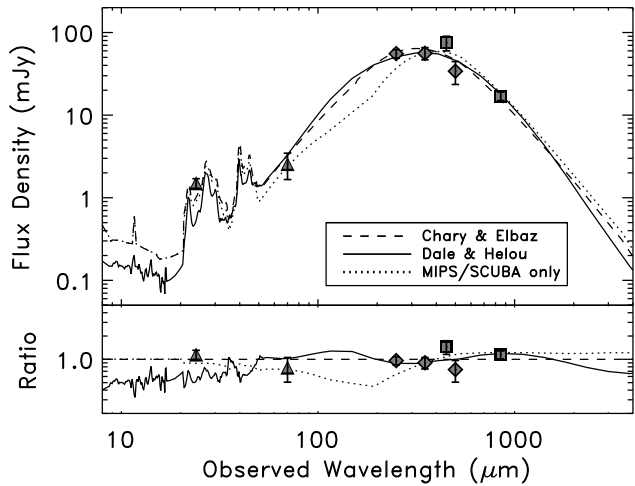}
\caption{The infrared SED of SMM J163554.2+661225.  The top panel shows flux densities from MIPS 24 and 70 $\mu$m (triangles), SPIRE 250, 350, and 500 $\mu$m (diamonds), and SCUBA 450 and 850 $\mu$m (squares) measured flux densities.  All error bars on the data points are shown at 1$\sigma$. The solid and dashed curves show the IR SED template fits to the 70, 250, 350, 450, 500, and 850 $\mu$m flux densities, using templates from DH02 (solid line) and CE01 (dashed line). The {\it Herschel} data constrains the peak of the dust emission to shorter wavelengths compared to the best-fit CE01 template to the SCUBA and MIPS 70 $\mu$m data alone, shown by the dotted line. The bottom panel shows the ratio of the best-fit models and data points to the best-fit model of the CE01 model (to all data points). As with the top panel, the error bars are 1$\sigma$.}
\end{figure}

\begin{figure}[th]
\epsscale{1.25}
\plotone{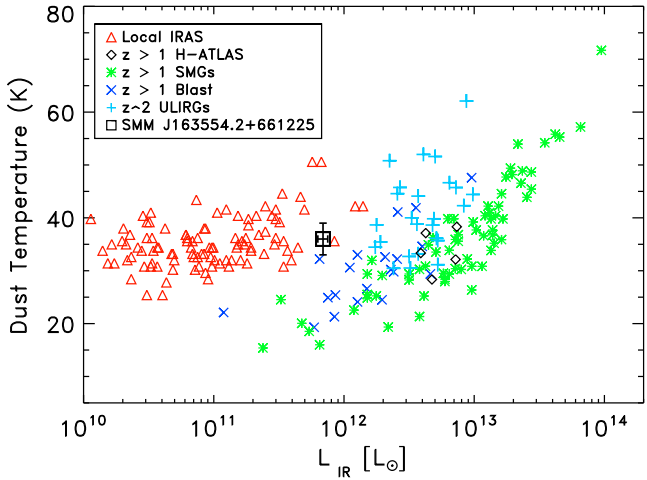}
\caption{Total IR luminosity versus dust temperature at z = 0 and z $\sim$ 2.  SMM J163554.2+661225 is shown by the black square. The error bars shown for SMM J163554.2+661225 are 1$\sigma$. Black diamonds are z $>$ 1 Herschel-ATLAS sources (Amblard et al.\ 2010); blue X's: z $>$ 1 BLAST sources (Dye et al.\ 2009); green asterisks: z $\sim$ 2 SMGs (Chapman et al.\ 2005); cyan crosses: z $\sim$ 2 ULIRGs with {\it Herschel} detections (Magdis et al.\ 2010); red triangles: local IRAS-selected galaxies with SCUBA 850$\mu$m detections (Dunne et al.\ 2000).}
\end{figure}

The CE01 SED templates were constructed to match correlations between the observed mid-IR and FIR fluxes of local galaxies. The shape of the FIR SED of the CE01 models are calibrated on the local IR-luminosity -- dust temperature relationship. As such, the CE01 templates with higher dust temperature correspond to galaxies with higher L$_{IR}$. However, we find that the shape of the FIR SED of SMM J163554.2+661225 corresponds to a much cooler dust temperature compared to local galaxies of similar luminosity.  Indeed, the CE01 template that best matches the FIR SED of SMM J163554.2+661225 has an intrinsic luminosity that must be ``scaled-up'' by a factor of 7.5. Therefore, this best-fit template must be scaled-up by a factor of nearly an order of magnitude to match both the shape of the SED and the flux density measurements of SMM J163554.2+661225. Similar results were seen by Muzzin et al.\ (2010) for their sample of two z $\sim$ 2 galaxies, where their galaxies were comparable to local luminous FIR galaxies, but scaled up in luminosity by more than an order of magnitude.  These new results confirm the conclusions of  Rigby et al.\ (2008), and P09 who used mid- and far-IR photometry and mid-IR spectroscopy to demonstrate that SMM J163554.2+661225 is inconsistent with the spectra of local ultra luminous infrared galaxies (ULIRGs), but is consistent with the SED shape of low-L$_{IR}$ local starburst galaxies and has just been ``scaled-up'' in luminosity and star formation by 1--2 orders of magnitude.  

Using the best-fit SEDs we calculated an effective dust temperature, T$_{D}$ = 36 $\pm$ 3 (1$\sigma$) K, assuming a modified grey-body distribution (e.g. Young et al.\ 1989; Calzetti et al.\ 2000; Amblard et al.\ 2010) with a fixed emissivity parameter, $\beta$=1.5, of the form:
\begin{equation}
f_{\nu} \propto \frac{\nu^{3+\beta}}{[\exp(\frac{h\nu}{kT_{dust}}) - 1]}
\end{equation}
The uncertainty on the dust temperature was derived by fitting all of the DH02 and CE01 models within 1$\sigma$ of the best-fit, based on the Monte Carlo simulations described above, to Equation 1.  Based on the mid-IR and SCUBA sub-mm data, P09 estimated T$_{D}$ = 52 K.   We favor the dust temperature measured here using the {\it Herschel} data, as these data now constrain the peak of the IR emission, providing the most constraining power on the dust temperature. The best-fit dust temperature again demonstrates that the SED shape of SMM J163554.2+661225 is not consistent with local galaxies of comparable L$_{IR}$, which have significant contributions of warm ($\gtrsim$ 70 K) dust to the IR emission (e.g., CE01; DH02; Rieke et al.\ 2009). 

We compare the calculated dust temperature and IR luminosity of SMM J163554.2+661225 to other galaxies that have measured dust temperatures using {\it Herschel} data or other sub-mm observations.  Figure 3 shows T$_{D}$ vs.\ L$_{IR}$ for galaxies at similar redshifts (1 $<$ z $<$ 3), including z $>$ 1 sources selected from the {\it Herschel}--ATLAS survey (Amblard et al.\ 2010), z $>$ 1 sources detected in BLAST (Dye et al.\ 2009), z $\sim$ 2 sub-millimeter galaxies (SMGs; Chapman et al.\ 2005), and {\it Herschel} detected z $\sim$ 2 ULIRGs (Magdis et al.\ 2010).  We also compare to a sample of local IRAS-selected galaxies with SCUBA 850$\mu$m detections (Dunne et al.\ 2000).  The other samples compute L$_{IR}$ (from 8-1000 $\mu$m) and T$_{D}$ in similar ways with a fixed $\beta$ = 1.5, except for the local IRAS sources which have varying values of $\beta$.  The IRAS sources also quote L$_{FIR}$ instead of L$_{IR}$, and we converted these using L$_{IR}$ = 1.4 $\times$ L$_{FIR}$ (Dale et al.\ 2001). If we change $\beta$ over a range of 2.0 to 1.2 for our object, then this corresponds to  a range in dust temperatures of 25 -- 42 K. However, if $\beta$ is allowed to vary, the best-fit value of $\beta$ = 1.6 results in T$_{D}$ = 34.5 K; consistent with our result for a fixed $\beta$ = 1.5.  

In Figure 3, SMM J163554.2+661225 has a similar dust temperature to the local IRAS sources, almost all of which have lower L$_{IR}$. It has a dust temperature that is lower than local IRAS sources of comparable L$_{IR}$. While SMM J163554.2+661225 has a somewhat warmer dust temperature than other z $\sim$ 1--2 SMGs of comparable L$_{IR}$; Chapman et al.\ (2005) discuss how the dust temperature of high-redshift sub-mm--selected samples are biased to cooler temperatures.  Furthermore, SMM J163554.2+661225 appears to to be in a ``bridge'' region between IR-luminous galaxies with cool and warm dust temperatures. Magdis et al.\ (2010) also note that their z $\sim$ 2 ULIRG sample with {\it Herschel} detections bridges the 'cooler' high-z SMGs to the 'warmer' local/intermediate-z ULIRGs. 

We estimate the total dust mass using our estimate of the dust temperature and the FIR luminosity. To estimate the total dust mass in SMM J163554.2+661225, we adopt the formulation of Young et al.\ (1989) and Calzetti et al.\ (2000):
\begin{equation}
M_{dust} = CS_{100}D^{2}[\exp(143.88/T_{dust}) - 1]~~~[M_{\odot}]
\end{equation}
Where S$_{100}$ is the rest-frame flux at 100 $\mu$m in Jy, D is the distance to the galaxy in Mpc, the expression in square brackets is the temperature-dependent part of the blackbody distribution, and C is a scale factor that depends on the physical properties of the dust grains: C $\sim$ 6 M\sol~Jy$^{-1}$ Mpc$^{-2}$ for $\beta$ = 2 and $\sim$5 M\sol~Jy$^{-1}$ Mpc$^{-2}$ for $\beta$ = 1.5 (Young et al.\ 1989; Calzetti et al.\ 2000).  For the range in single component dust temperatures, we estimate a total dust mass for this galaxy of 2.8 $^{+1.2}_{-0.8}$ (1$\sigma$) $\times$ 10$^{8}$ M\sol, corrected for the gravitational-lensing magnification factor of 22. This is approximately a factor of 10 higher than the estimated dust mass from Kneib et al.\ (2005) based on the 850 $\mu$m flux density and a dust temperature of $\sim$ 50 K. This can be explained by the colder dust temperature determined here; as there will be a larger fraction of cold dust for this galaxy with a dust temperature of 36 K compared to 50 K, resulting in a higher total dust mass.  Based on the molecular gas estimate of 4.5$\times$10$^{9}$ M\sol~by Kneib et al.\ (2005) this implies a dust--to--gas mass ratio of 0.06.  This is consistent with ratios found for local star forming IR galaxies in the Survey of Nearby Infrared Galaxies (SINGS; Kennicutt et al.\ 2003; Draine et al.\ 2007) sample.

P09 also noted that SMM J163554.2+661225 has a lower ratio of L(24 $\mu$m)/L(Pa$\alpha$) compared to what is measured for any local ULIRGs, demonstrating that SMM  J163554.2+661225 lacks a warm (T$_{D}$ $\sim$ 70 K) dust component.  However, there could be a colder dust component present in SMM J163554.2+661225.  The local samples of IRAS selected galaxies (Dunne et al.\ 2000) preferentially detected galaxies with large amounts of T$_{D}$ $>$ 40 K, but because these galaxies were selected with IRAS they miss most of the cold dust dominated sources. The observations of optically selected galaxies of the SCUBA Local Universe Galaxy Survey (SLUGS; Vlahakis et al.\ 2005), supplemented by {\it Spitzer} data of a subsample of SLUGS galaxies by Willmer et al.\ (2009), showed that the mid--to--far-IR SEDs of these galaxies are better matched with both cold and warm dust components. These samples show on average a 2$\times$ larger proportion of cold dust relative to warm dust than solely IR-selected galaxy samples do, such as the IRAS selected sample (Dunne et al.\ 2000) and SINGS sample (Kennicutt et al.\ 2003).  

We fit models of Draine \& Li (2007) to the SED of SMM J163554.2+661225 in order to investigate a two component dust fit.  The implied total IR luminosity from the best-fit Draine \& Li model is 7.0$\times$10$^{11}$ L\sol, consistent with our other SED fits. The best-fit Draine \& Li (2007) SED is modeled with a modified grey-body distribution with two dust components, each with $\beta$=1.5. This fit corresponds to warm and cold effective dust temperatures of 46 K and 28 K.  Using the dust mass equation above, we calculate warm and cold dust mass components of M$_{cold}$ = 4.3$\times$10$^{8}$ M\sol, and M$_{warm}$ = 5.6$\times$10$^{7}$ M\sol. The mass from the warm dust component more closely matches the estimated total dust mass determined by Kneib et al.\ (2005).  The ratio of M$_{cold}$/M$_{warm}$ = 7.8, with a total dust mass = 4.9$\times$10$^{8}$ M\sol, consistent within 2$\sigma$ of our result of the dust mass estimate from the single-component temperature fits. This lower ratio of cold dust to warm dust for SMM J163554.2+661225 is more similar to that of the SINGS sample than in the SLUGS sample (Draine et al.\ 2007; Willmer et al.\ 2009).  The SLUGS sample typically had higher ratios of M$_{cold}$/M$_{warm}$, on order $\sim$1000 (Willmer et al.\ 2009).  This agrees with the result of Rigby et al.\ (2008) who noted the similarity between the mid-IR emission features of SMM J163554.2+661225 and that of local IR star-forming galaxies.

\section{Star Formation Rates and Implications for IMF} 

Local star-forming galaxies show a tight trend between L(Pa$\alpha$)
and L$_{IR}$ (e.g., Calzetti et al.\ 2005; Alonso-Herrero et al.\
2006), both of which trace the instantaneous SFR.  We calculate the
SFR of SMM J163554.2+661225 based on the derived total IR luminosities
from the SED fitting, using the relation derived by Kennicutt (1998),
which gives SFR = 4.5$\times$10$^{-44}$ $\times$ FIR (erg s$^{-1}$) =
119 $\pm$ 10 M\sol~yr$^{-1}$ for a Salpeter IMF from 0.1 -- 100 M\sol.
Previous estimates of SFRs from the total IR emission in this galaxy range from
SFR$_{IR}$ = 90 -- 180 M\sol~yr$^{-1}$ (P09) and SFR$_{IR}$ = 140
$\pm$ 30 M\sol~yr$^{-1}$ (Rigby et al.\ 2008).    Our estimate of the
IR emission from \textit{Spitzer}, \textit{Herschel}, and the sub-mm
are consistent with these results, although the uncertainty on our
measurement is significantly reduced by including the
\textit{Herschel} far-IR data.    

P09 obtained a Pa$\alpha$ measurement of component B using {\it
Spitzer} IRS spectroscopy with the SL2 module and measured a
Pa$\alpha$-SFR; the {\it Spitzer} IRS observations only covered
component B, so we can only compare the IR-derived SFR to the
Pa$\alpha$-SFR for component B.  P09 measured a Pa$\alpha$ line flux
for component B of 8.6 $\pm$ 1.4 $\times$ 10$^{-16}$ erg s$^{-1}$
cm$^{-2}$, which is uncorrected for dust attenuation and
gravitational-lensing magnification.  Kneib et al.\ (2004) measured an
H$\alpha$ line flux of 5.9 $\times$ 10$^{-16}$ erg s$^{-1}$ cm$^{-2}$.
We re-calculate the dust-corrected Pa$\alpha$ luminosity, using an updated value for the dust attenuation for the Pa$\alpha$
line (see P09).  We use the ratio of the Pa$\alpha$ line flux to the
H$\alpha$ line flux to estimate the amount of dust attenuation
affecting the nebular gas, assuming the Calzetti et al.\ (2000) dust
law.  This results in an extinction estimate of A(V) = 3.9 $\pm$ 0.4
mag, and this corresponds to A(Pa$\alpha$) = 0.58 $\pm$ 0.04 mag (we assume that the stellar continuum is attenuated at the same level as the nebular lines [e.g. Erb et al.\ 2006; though see also F{\"o}rster Schreiber et al.\ 2009]). Correcting for the dust extinction, and applying a gravitational-lensing magnification factor of 22, the Pa$\alpha$ line
luminosity is L(Pa$\alpha$)$_{cor}$ = 3.38 $\pm$ 0.55 $\times$
10$^{42}$ erg s$^{-1}$.  This corresponds to an extinction and
lensing--corrected  SFR$_{Pa\alpha}$ = 225 $\pm$ 37 M\sol~yr$^{-1}$,
following the Kennicutt (1998) relations. This luminosity corresponds
to an ionizing continuum flux of Q$_{ion}$ = 2.1 $\pm$ 0.3 $\times$
10$^{55} \gamma$ s$^{-1}$.  This updated dust correction results in a 30$\%$ increase in the
derived Pa$\alpha$-luminosity and Pa$\alpha$-SFR, and Q$_{ion}$ from
that originally derived by P09.

Our estimate of the SFR derived from the IR emission, SFR(IR) =
$119\pm 10$ M\sol~yr$^{-1}$ is significantly different than the SFR
calculated from the dust--corrected Pa$\alpha$ line, SFR(Pa$\alpha$) =
225 $\pm$ 37 M\sol~yr$^{-1}$.    They are offset by a factor of 1.9
$\pm$ 0.35, a 2.5$\sigma$ discrepancy (or a 99$\%$ confidence).   The offset between the SFR
derived from the Pa$\alpha$ line and IR implies that one or more of
our assumptions are incorrect.  We consider here several explanations
to account for the apparent discrepancy between the IR and Pa$\alpha$
SFRs.  

\subsection{Variations in Dust Attenuation}

In our analysis, we corrected the Pa$\alpha$ line luminosity assuming
a Calzetti dust extinction law (Calzetti et al.\ 2000).     The exact
level of dust attenuation is uncertain (see below), but changes in
the assumed extinction are  unable to account fully for the  measured
offset between the IR and Pa$\alpha$ SFRs.  For example,   P09 report
a Pa$\alpha$ luminosity of $(2.05\pm 0.33) \times 10^{42}$ erg
s$^{-1}$ \textit{uncorrected} for dust extinction.  This corresponds
to a SFR of $130 \pm 21$ M$_\odot$ yr$^{-1}$.   If there is no dust extinction correction the SFRs are consistent at the $1\sigma$ level.   However, zero dust attenuation seems unlikely.
First, the ratio of the measured nebular emission lines are strongly
inconsistent with the assumption of zero dust (see below).  Second,
P09 showed from modeling the rest-frame UV-to-NIR photometry of the
galaxy with stellar population synthesis models (Bruzual \& Charlot
2003) that the dominant star-forming component has a best fit with
A(V)  $\sim$ 3.2 mag.  Finally, the simple fact that this galaxy is bright in the far-infrared implies that it contains a significant amount of dust (see \S 3.1).  Therefore, significant dust extinction seems
required. 

In our analysis, using the dust extinction law of Calzetti et
al. (2000) results in a line extinction measurement of  A(Pa$\alpha$)
= 0.58 mag derived from the observed ratio of the H$\alpha$ and
Pa$\alpha$ lines.   Using other dust attenuation laws (e.g., Dopita et
al.\ 2005; Cardelli et al.\ 1989) results in slightly larger
extinction estimates, and would therefore \textit{increase} the
dust-corrected Pa$\alpha$ luminosity by $\sim$10\%.  This amplifies
the offset between the  Pa$\alpha$-derived SFR and IR-derived SFRs.
Thus, the use of the Calzetti dust law results in a conservative
estimate of the Pa$\alpha$ luminosity and SFR$_{Pa\alpha}$, and
therefore minimizes the differences between the two SFR measurements. 

The Pa$\alpha$ dust extinction determined above is based on the ratio
of the H$\alpha$ line flux from Kneib et al.\ (2004) to the Pa$\alpha$
line flux.  However, there are some discrepancies in the reported
H$\alpha$ line flux measurement, and it could be as much as 4$\times$
fainter (P09, W. Rujopakarn et al., in preparation), but a fainter
H$\alpha$ line flux would image a \textit{larger} derived A(V) and
A(Pa$\alpha$), and would increase the extinction-corrected Pa$\alpha$
luminosity and SFR (e.g., a H$\alpha$ line flux lower by a factor of 4
would increase dust-corrected Pa$\alpha$ line flux by $\sim$ 30$\%$). 

We also we re-calculate A(Pa$\alpha$) using a newer measurement of
H$\beta$ for this galaxy from Richard et al.\ (2011).  The observed
Pa$\alpha$ line flux for this galaxy again is 8.6 $\pm$ 1.4 $\times
10^{-16}$ erg s$^{-1}$, uncorrected for the gravitational lensing
magnification or dust attenuation (P09).  The observed H$\alpha$ line
flux for component B (making no correction for uncertainties in
gravitational lensing, dust attenuation, or slit losses) is 5.9
$\times 10^{-16}$ erg s$^{-1}$ (Kneib et al.\ 2004), and the H$\beta$
line flux is 6.88 $\pm$ 0.29 $\times 10^{-17}$ erg s$^{-1}$ (Richard
et al.\ 2011).  Given these three lines we calculate an average
E(B-V) of 0.97 $\pm$ 0.07 and this corresponds to an
average A(Pa$\alpha$) of 0.57 $\pm$ 0.06 assuming the Calzetti et al.\
(2000) attenuation law. Therefore, the addition of the H$\beta$ line confirms the estimate of the dust attenuation affecting the nebular
gas, and implies the extinction is approximately optically thin to
these photons. 

Lastly,  the extinction correction derived here assumes the nebular
gas is approximately optically thin to photons from H$\alpha$ and
Pa$\alpha$. While this is supported by the measured line ratios, if
the opacity of the nebular gas is substantially higher, then it will
obscure a higher fraction of emission associated with star formation.
However, if this is the case, then the A(Pa$\alpha$) we derive above
is only lower limit, as it is only sensitive to the outermost layer of
the gas.   If this is the case, then the dust corrected Pa$\alpha$
luminosity will also be  \textit{higher}, which further exacerbates
the problem.   

Therefore, in summary it seems unlikely that assumptions about the
dust law or the measurements of the dust attenuation of the Pa$\alpha$
line contribute significantly to the observed offset between the IR
and Pa$\alpha$ SFRs.   Indeed, our assumptions about the dust
attenuation are mostly conservative and the intrinsic Pa$\alpha$ may
be larger than reported here. 

\subsection{Variations to the ISM Conditions}

The calculations above depend on  intrinsic line ratios assuming Case
B recombination at T$_{e}$ = 10$^{4}$ K.  We also consider how changes
in the assumed ISM temperatures effect the estimation of
A(Pa$\alpha$), and therefore the derived Pa$\alpha$
luminosity. Increasing the ISM temperature to larger values (T$_{e}$ =
2$\times$10$^{4}$ K), as could be expected for a highly star-forming
galaxy, increases the intrinsic value of H$\alpha$ / Pa$\alpha$ from
8.46 to 9.68 (Osterbrock 1989). This  increases the estimated value of
A(Pa$\alpha$) to 0.62 mag, an increase of 10\%.  This would slightly
increase the dust-corrected values of L(Pa$\alpha$) and
SFR$_{Pa\alpha}$.  

As an extreme case, A(Pa$\alpha$) would be  reduced to 0.46 mag if one
assumes Case A recombination and a very low ISM temperature of T$_{e}$
= 2500 K.  This would decrease L(Pa$\alpha$)$_{cor}$ and
SFR$_{Pa\alpha}$ by a maximum of 12$\%$, but it is insufficient to
account for the factor of $\sim$2 difference between the Pa$\alpha$
and IR SFRs.  Therefore, our
assumption of Case B recombination with an ISM temperature of 10$^{4}$
K does not affect our conclusions.  

\subsection{Changes to the Star Formation History}

The offset between the two SFRs could also be due to assumptions about
the stellar population of SMM J163554.2+661225. The Pa$\alpha$ and
L$_{IR}$ SFR calibrations both assume a constant SFR and a stellar
population age of 100 Myr, as well as a Salpeter IMF.  Using the 2007
version of the Bruzual \& Charlot (2003) stellar population synthesis
models, we investigated if this factor of 1.9 between the derived SFRs
could be accounted for by varying the star formation history or age,
but still with a Salpeter IMF.     For example, the commonly assumed
relations of Kennicutt 1998 assume a roughly constant star--formation
history for the past 100 Myr.  However, there is evidence that
galaxies at $z > 2$ have star-formation histories that increase with
time (e.g., Papovich et al.\ 2011), which affects the relative
proportion of stars that contribute to nebular emission and the IR
emission.  

From the stellar population synthesis models with a Salpeter IMF, we
calculate the ratio of ionizing photons (Q$_{ion}$) to bolometric
luminosity versus stellar population age. We assume that the ratio of
Q$_{ion}$ / L$_{bol}$ is proportional to the ratio of Pa$\alpha$
luminosity to total IR luminosity (as argued by Kennicutt 1998).  We
tested how a SFR that  rises with time affects the ratio of Q$_{ion}$
/ L$_{bol}$.   Assuming the SFR rises with time (approximated by the delayed SFR model within Bruzual \& Charlot 2003),  we
find a maximal increase  of  Q$_{ion}$ / L$_{bol}$ by at most a factor
of 1.25 compared to this ratio for constant SFR. Other increasing SFRs (including a model with exponentially increasing SFR) were also investigated which produce more Lyman-continuum photons by as much as a factor of 1.3 for ages $\lesssim$ a few e-folding times.  Higher ratios are possible, but only in extreme situations, such as exponentially increasing SFRs at ages $>>$ a few e-folding times, but it seems physically unlikely that a galaxy will increase its SFR exponentially for such sustained periods without disruption (i.e.\ feedback). Therefore, star formation history alone can only partially explain the offset between
the two derived SFRs. 

\subsection{Varying the IMF}

One possible way to create more ionizing photons would be an IMF that
is weighted towards high-mass stars compared to that of an IMF with a
Salpeter-like IMF slope.    We investigated how changes in the the
form of the IMF affects the Q$_{ion}$ / L$_{bol}$ ratio, including an
IMF with a flatter slope (i.e., a ``top--heavy" IMF) or an IMF with a
higher upper mass cutoff.  The expected ratio of the total number of
ionizing photons (Q$_{ion}$) to the total IR luminosity, Q$_{ion}$ /
L$_{IR}$, following the Kennicutt (1998) relations and assuming a
Salpeter IMF from 0.1 to 100 M\sol, should have a ratio of
1.6$\times$10$^{43}$ $\gamma$ s$^{-1}$ L\sol$^{-1}$.  This assumes
that the Pa$\alpha$-derived SFR and the IR-derived SFR should be
equal.  However, our measured ratio for SMM J163554.2+661225 is
Q$_{ion}$ / L$_{IR}$ is 3.0 $\pm$ 0.6 $\times$10$^{43}$ $\gamma$
s$^{-1}$ L\sol$^{-1}$, reflecting the factor of 1.9 difference between
the two SFRs compared to the theoretical value.

We computed Q$_{ion}$ and L$_{IR}$ (assumed to be $\sim$ L$_{bol}$) as
a function of both the IMF slope and upper-mass cutoff.  To make this
calculation, we take empirical measures of Q$_{\mathrm{0}}$  as a
function of spectral type for O-stars with masses 19.3 -- 100
M\sol~derived by Sternberg et al.\ (2003)\footnotemark. We then
weighted the values by the number of stars per unit mass for each IMF,
and integrate to calculate the total number of ionizing photons for a
given IMF.  We repeated the calculation for the bolometric luminosity,
assuming a mass--luminosity relation of L $\propto$ M$^{3.8}$ for
stars less than 20 M\sol~(Popper 1980), and L $\propto$ M$^{1.9}$ for
M $>$ 20 M\sol~based on empirical models of OB stars (Sternberg et
al.\ 2003).  

\footnotetext[3] {We assume that Q$_{\mathrm{0}}$ increases with stellar mass at a rate derived from the empirical measurements of Sternberg et al.\ (2003). We extrapolated this rate to estimate Q$_{\mathrm{0}}$ values above 100 M\sol.}

\begin{figure*}[th]
\centering
\epsscale{1.0}
\plotone{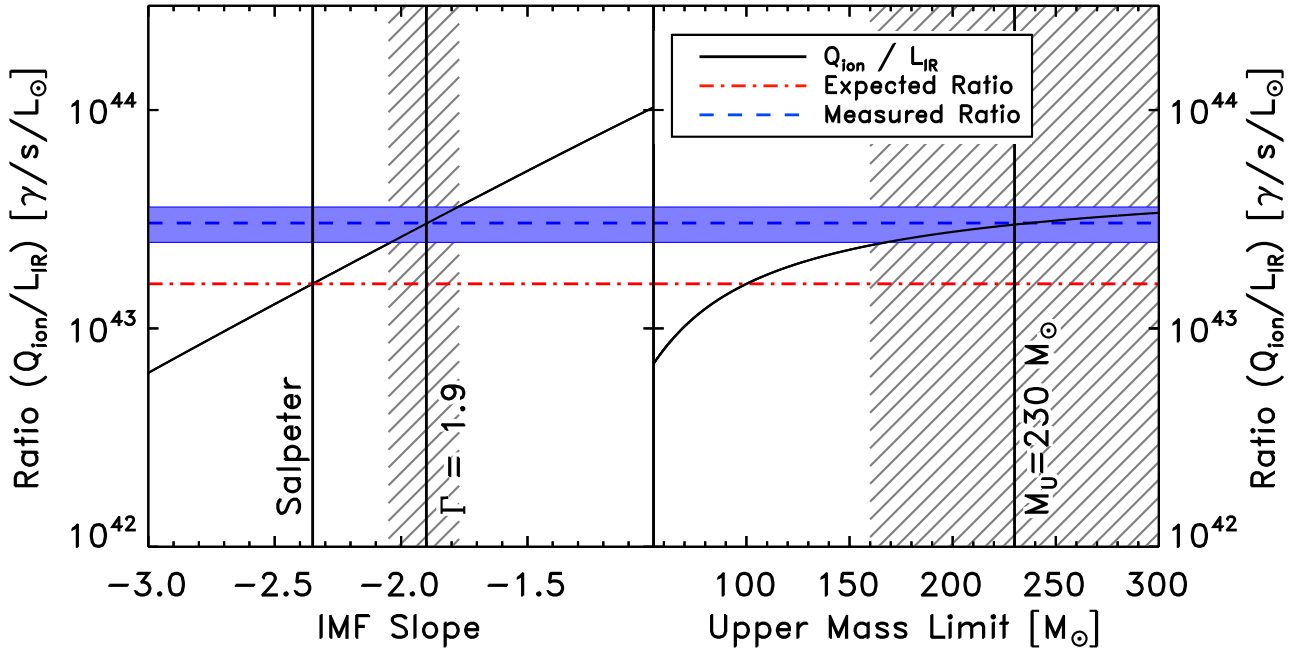}
\caption{Effect of changes to the IMF on the ratio of the number of ionizing photons (Q$_{ion}$) to  L$_{IR}$.  Both scenarios or a combination can reproduce the measured difference between the SFR$_{IR}$ and SFR$_{Pa\alpha}$. The left plot shows how varying the IMF slope changes the derived ratio of Q$_{ion}$ to L$_{IR}$. The red dot-dash line shows the expected ratio following the Kennicutt relations assuming a Salpeter mass distribution with upper and lower mass cutoffs of 100 M\sol~and 0.1 M\sol. The blue dashed line shows our measured value of Q$_{ion}$ / L$_{IR}$, the blue shaded area shows the 1$\sigma$ uncertainty on the measured ratio.  To match our observations, more ionizing photons need to be made relative to the L$_{IR}$ (traced by the bolometric luminosity, see Kennicutt 1998) which can happen with a shallower IMF distribution with slope $\Gamma$ = 1.9 $\pm$ 0.15.  The right plot shows that we can also reproduce our observed ratio of Q$_{ion}$ / L$_{IR}$ with a Salpeter IMF by extending the upper-mass limit to 200 M\sol. The grey hatched region in each panel shows our preferred 1$\sigma$ values.}
\vspace{2mm}
\end{figure*}  

Figure 4 shows the effect on the ratio Q$_{ion}$ / L$_{IR}$ for IMFs
of different forms, including changes in the IMF slope and the
upper-mass cutoff.  We have overplotted both the expected ratio from
Kennicutt (1998) and our measured ratio.  The left panel shows that a
shallower IMF slope, with a best-fit of $\Gamma$ = 1.9 $\pm$ 0.15,
reproduces our measured ratio of Q$_{ion}$ / L$_{IR}$.  This result is
a 3$\sigma$ variation from Salpeter ($\Gamma = 2.35$). As shown in the
right panel, a Salpeter IMF slope with an upper-mass cutoff extended
up to $\sim$200 M\sol~also reproduces the measured ratio. The
1$\sigma$ uncertainty on the measured ratio of Q$_{ion}$ / L$_{IR}$ is
also shown in Figure 4 by the blue shaded area.  The vertical hatched
region in each panel shows our preferred 1$\sigma$ values. Therefore,
a flattening of the IMF slope to $\Gamma$ = 1.9 $\pm$ 0.15 or an
increase in the IMF upper-mass cutoff to at least 160 M\sol\ (or a
combination of both) is able to account for the observed Pa$\alpha$
line luminosity and IR luminosity. 

P09 estimate that SMM J163554.2+661225 has a total stellar mass of
$\sim$ 10$^{10}$ M\sol. This implies that the total number of O stars
(M $\geq$ 20 M\sol) is about 10$^{7}$, assuming a Salpeter IMF from
0.1 to 100 M\sol~and a constant SFR. If the IMF slope varies from the
nominal value of Salpeter to $\Gamma$ = 1.9 then the number of O stars
between 20 --100 M\sol~would increase by a factor of 3.5.
Alternatively, this additional number of O stars can be accounted by
extending the upper mass cut off to at least 160 M$_\odot$, this would result in $\sim$5$\%$ more O stars above 100 M\sol, in which
case the majority of the additional ionization comes from these
highest mass stars.

\subsection{Metallicity Effects}

We also investigate if changes in metallicity can account for the
measured discrepancy in the SFRs.  In the above calculations for a
varying SFH or a changing IMF, we assume solar metallicity. Using
stellar population synthesis models from Starburst 99 (Leitherer et
al.\ 1999), we specifically try to determine if lower metallicities
can reproduce the observed ratio of Q$_{ion}$ / L$_{bol}$.     We ran
four simulations in Starburst 99, only varying the metallicity.  In
all four simulations, we assumed a constant SFR of 120
M\sol~yr$^{-1}$, and a Salpeter IMF between 0.1 -- 100 M\sol.  The
four simulations had metallicities of 0.02 Z\sol, 0.2 Z\sol, 0.4
Z\sol, and solar metallicity.  Based on these results, we found that
the simulation with solar metallicity at an age of 100 Myr reproduces
the expected ratio of the number ionizing photons to L$_{bol}$
(1.6$\times$10$^{43}$ ionizing $\gamma$ s$^{-1}$ L\sol$^{-1}$) as
calculated following the Kennicutt (1998) relations as described
above.  This is the factor of 1.9 lower than our measured ratio of
Q$_{ion}$ / L$_{bol}$.  In order to reproduce our measured ratio of
Q$_{ion}$ / L$_{bol}$, the metallicity must be lowered to $\lesssim$
0.02 Z\sol.   Therefore, a very low metallicity population could
reproduce the higher rate of ionizing photons as compared to the total
luminosity without changing the IMF. 

However, we have reason to believe that the metallicity of this galaxy
is nearly solar.  Using the line ratios of [N II]/H$\alpha$ = 0.3
$\pm$ 0.1 (Kneib et al.\ 2004) and [O III] /H$\beta$ = 1.5 $\pm$ 0.11
(Richard et al.\ 2011) we calculate the metallicity based on the N2
and O3N2 indices.  From the N2 index, we get a value of 12 + log(O/H)
= 8.6 $\pm$ 0.1, assuming the relation from Pettini \& Pagel (2004),
consistent with the solar value 12 + log(O/H) = 8.66.  From the O3N2
index, we get a value of 8.5 $\pm$ 0.06, again close to solar
metallicity. Therefore the gas metallicity of SMM J163554.2+661225
appears to be consistent with solar metallicity, and is inconsistent
with a metallicity of 0.02 Z\sol~by more than 4$\sigma$.  Since they have just formed, we can also
expect the massive OB stars that produce the increased fraction of
ionizing photons to have a similar metallicity as the gas, and
therefore it is not likely that a very low metallicity stellar
population is responsible for the high measured value of Q$_{ion}$ /
L$_{bol}$.

\subsection{Differential Magnification}

As a final possibility for a cause of the differing SFRs we look at
the lensing model of SMM J163554.2+661225 in order to determine if
there is evidence for differential magnification between the physical
region of the galaxy that produces the Pa$\alpha$ emission relative to
the IR producing regions.   For example, if Pa$\alpha$ and IR emitting
regions have different gravitational-lensing magnification factors,
then it could contribute to the discrepancies we observe.   

The gravitational lensing model at the astrometric position of the
galaxy shows variations in magnification of order 10-20\% over the 1
arcsecond scale of the galaxy (J.-P. Kneib, private communication,
2011), and this could contribute to the measured offset.
However,  it seems more likely that the Pa$\alpha$ and IR emitting
regions are coaligned.    We note that the \textit{Spitzer/IRAC} data
and MIPS 24~\micron\ centroids are aligned to better than 1 arcsecond,
suggesting the  regions contributing to the emission at these
wavelengths are magnified equally.   Therefore, while we can not
exclude that variations in gravitational lensing contribute, we argue
that it is unlikely to dominate the measured offset. 

\section{Summary and Final Thoughts}

The {\it Herschel} SPIRE data of SMM J163554.2+661225 allow for a highly accurate fit to the infrared SED of this z=2.515 lensed galaxy.  We measured a best-fit IR luminosity of 6.9 $\pm$ 0.6 $\times$10$^{11}$ L\sol, corresponding to a dust temperature of 36 $\pm$ 3 K, and a dust mass of $\sim$3$\times$10$^{8}$ M\sol.  The  IR-derived SFR is 119 $\pm$ 10 M\sol~yr$^{-1}$ which is a factor of 1.9 lower than the previously derived Pa$\alpha$-SFR.  In order to account for the discrepancy in the derived SFRs, we have analyzed several different scenarios that could alter the ratio of Pa$\alpha$ luminosity and L$_{IR}$.  One possibility is that there is a varying SFH due to a rising SFR.  Using a SFR that increases with time, with a Salpeter IMF can increase the number of ionizing photons relative to the total bolometric luminosity (and thereby to the L$_{IR}$), by as much as 25-30$\%$ at a given stellar population age, but cannot by itself account for the measured difference between SFR$_{IR}$ and SFR$_{Pa\alpha}$.  

Varying the high-mass end of the IMF can explain our results, either
by increasing the high-mass cutoff up to at least 160 M\sol, or
changing the IMF slope to $\Gamma = 1.9$ from the  Salpeter value of
$\Gamma = 2.35$, or a combination of the two.  Either scenario
produces an increase in the number of massive stars, resulting in
higher rates of ionizing photons. This appears necessary to explain
the higher Pa$\alpha$-derived SFR over the IR-derived SFR.  However,
with current data we are unable to distinguish between these two
scenarios.  Similar changes in the IMF at the high mass end have been
shown to have an effect on the relative contributions between various
SFR indicators (e.g., Daddi et al.\ 2007), whereas extending the IMF
upper-mass limit up to at least 120 M\sol~has also been shown to be
important in reproducing the observed H$\alpha$ flux in star-forming
galaxies from the SDSS (Hoversten \& Glazebrook 2008; 2010).  Based on
our analysis, we favor a scenario where the IMF may require
minor adjustments to the standard form of the Salpeter IMF for some
star-forming galaxies.

Flattening of the IMF at higher redshifts has been proposed by other
studies to account for offsets between stellar mass and SFR densities
(e.g., Dav\'e 2008 and references therein). A flat IMF, with $\Gamma$
= 1, has also been proposed to account for SMG number counts (Baugh et
al.\ 2005; Lacey et al.\ 2008).   Our results of SMM  J163554.2+661225
allow for an IMF slope $\Gamma$ = 1.9 $\pm$ 0.15, only slightly
flatter than Salpeter. Additionally, it could be that both a slight
change in the IMF and a change to the SFH are at play. If the SFH is
changing or there has been a burst in star formation activity, this
could mimic some of the changes produced via a varying IMF.  However,
a more complex star formation history only cannot explain the whole
effect we see; some alteration of the IMF is still needed.  As discussed above, if the galaxy has a SFR that is increasing with time, then
this would increase the expected number of ionization photons by
25$\%$ compared to a constant SFR.  In this case, coupled with the increasing SFR then the IMF would only
be slightly flatter than Salpeter with a slope $\Gamma$ $\sim$ 2.06
and/or with an upper-mass cutoff of  up to 120 M\sol. 

We also investigated if changes to the assumed metallicity and dust
content of this galaxy could account for the offset in derived SFRs.
This galaxy would need a very low metallicity population of $\sim$
0.02 Z\sol, however this low metallicity does not match with the
observed line ratios of the gas, which give a measured metallicity of
$\sim$ solar.   A lower dust extinction of A(V) = 1.0 mag,
corresponding to E(B-V) = 0.25 mag, and A(Pa$\alpha$) = 0.15 mag,
could also account for the discrepancy between the IR-derived and
Pa$\alpha$-derived SFRs. However changing the assumptions about the
ISM conditions (ie. Temperature or Case A vs.\ Case B recombination)
only have a small effect on the derived A(V) values for this galaxy,
and at most can only decrease the derived Pa$\alpha$-SFR by 12$\%$,
again not enough to make up the difference between the two SFRs.  Also
from stellar population synthesis modeling, previous authors (P09)
have shown that this galaxy is best-fit by a two-component stellar
population fit, with the dominant star-forming component being very
dusty, corresponding to E(B-V) = 0.8 mag and A(V) $\sim$ 3.2, and
therefore inconsistent with A(V) $\sim$ 1. We therefore conclude that
the assumptions we have made about the metallicity and dust content of
this galaxy are consistent with the known observations of this galaxy,
and it would take extreme changes to one or the other in order to be
able to explain the differences in the derived SFRs.  

Another possibility to explain the offset is that the regions of the
galaxy that emit the Pa$\alpha$ and IR may experience different
amounts of gravitational lensing magnification.  However, we expect
the Pa$\alpha$ emission to trace the star-forming and IR dominated
regions.  Given the size of the galaxy, this effect seems to be less
than $\approx 20$\%, and likely is not a dominant effect. 

As a final thought, it may be that a combination of all the effects
(variations in the star-formation history, minor adjustments to the
IMF, variations in extinction, etc.) may all contribute to the
difference in the Pa$\alpha$-derived SFR and the IR SFR.    Future
observations of z$\sim$2 lensed galaxies that have both Pa$\alpha$
emission and FIR observations may be able to determine if this offset
is common at z$\sim$2 or if SMM J163554.2+661225 is an unusual case.

\acknowledgements   The authors wish to thank Kim-Vy Tran for many
useful conversations, Jean--Paul Kneib for his help on questions regarding the lensing model, as well as Rob Ivison, and
Anthony Smith for help with questions regarding the HerMES survey. We
also thank the anonymous referee for a very useful report which
improved the quality of this paper. This research has made use of data
from HerMES project (Oliver et al.\ 2010;
http://hermes.sussex.ac.uk/). HerMES is a Herschel Key Programme
utilizing Guaranteed Time from the SPIRE instrument team, ESAC
scientists and a mission scientist. The HerMES data was accessed
through the HeDaM database (http://hedam.oamp.fr) operated by CeSAM
and hosted by the Laboratoire d'Astrophysique de Marseille.  This work
is based in part on observations made with the {\it Herschel Space
Observatory} and the \textit{Spitzer Space Telescope}, which is
operated by the Jet Propulsion Laboratory, California Institute of
Technology under a contract with NASA. Support for this work was
provided by NASA through an award issued by JPL/Caltech. Further
support for KDF, CP, and SLF was provided by Texas A\&M University. SLF also received support by NASA through Hubble Fellowship grant HST-HF-51288.01, awarded by the Space Telescope Science Institute, which is operated by the Association of Universities for Research in Astronomy, Inc., for NASA, under contract NAS 5-26555.


\begin{thebibliography}{}
\bibitem[Alonso-Herrero et al.\ (2006)]{alo06} Alonso-Herrero, A., Rieke, G. H., Rieke, M. J., Colina, L., P\'erez-Gonz\'alez, P. G., \& Ryder, S. D. 2006, ApJ, 650, 835
\bibitem[Amblard et al.\ (2010)]{amb10} Amblard, A., et al.\ 2010, A\&A, 518, L9
\bibitem[Baldry \& Glazebrook (2003)]{bg03} Baldry, I. K. \& Glazebrook, K. 2003, ApJ, 593, 258
\bibitem[Baugh et al.\ (2005)]{bau05} Baugh, C. M., et al.\ 2005, MNRAS, 356, 1191
\bibitem[Bertin \& Arnouts (1996)]{ba96} Bertin, E. \& Arnouts, S. 1996, ApJS, 117, 393
\bibitem[Bruzual \& Charlot (2003)]{bc03} Bruzual, G. \& Charlot, S. 2003, MNRAS,  344, 1000
\bibitem[Calzetti et al.\ (2005)]{cal05} Calzetti, D., et al.\ 2005, ApJ, 633, 871
\bibitem[Calzetti et al.\ (2000)]{cal00} Calzetti, D., Armus, L., Bohlin, R. C., Kinney, A. L., Koornneef, J., \& Storchi-Bergmann, T. 2000, ApJ, 533, 682
\bibitem[Cardelli et al.\ (1989)]{car89} Cardelli, J. A., Clayton, G. C., \& Mathis, J. S. 1989, ApJ, 345, 245
\bibitem[Chapman et al.\ (2005)]{cha05} Chapman, S. C., Blain, A. W., Smail, I., \& Ivison, R. J. 2005, ApJ, 622, 772
\bibitem[Chapman et al.\ (2002)]{cha02} Chapman, S. C., Scott, D., Borys, C., Fahlman, G. G., 2002, MNRAS, 330, 92
\bibitem[Chary \& Elbaz (2001)]{ce01} Chary, R. \& Elbaz, D. 2001, ApJ, 556, 562 (CE01)
\bibitem[Daddi et al.\ (2007)]{dad07} Daddi, E. et al.\ 2007, ApJ, 670, 156
\bibitem[Dale \& Helou (2002)]{dh02} Dale, D. A. \& Helou, G. 2002, ApJ, 576, 159 (DH02)
\bibitem[Dale et al.\ (2001)]{dal01} Dale, D. A., Helou, G., Contursi, A., Silbermann, N. A., \& Kolhatkar, S. 2001, ApJ, 549, 215
\bibitem[Dav\'e (2008)]{dav08} Dav\'e, R. 2008, MNRAS, 385, 147
\bibitem[Dopita et al.\ (2005)]{dop05} Dopita, M. A., et al.\ 2005, ApJ, 619, 755
\bibitem[Draine \& Li (2007)]{dl07} Draine, B. T. \& Li, A. 2007, ApJ, 657, 810
\bibitem[Draine et al.\ (2007)]{dra07} Draine, B. T., et al.\ 2007, ApJ, 663, 866
\bibitem[Dunne et al.\ (2000)]{dun00} Dunne, L., Eales, S., Edmunds, M., Ivison, R., Alexander, P., \& Clements, D. L. 2000, MNRAS, 315, 115
\bibitem[Dye et al.\ (2009)]{dye09} Dye, S., et al.\ 2009, ApJ, 703, 285
\bibitem[Elbaz et al.\ (2010)]{elb10} Elbaz, D., et al.\ 2010, A\&A, 518, L29
\bibitem[Elmegreen (2006)]{elm06} Elmegreen, B. G. 2006, ApJ, 648, 572
\bibitem[Erb et al.\ (2006)]{erb06} Erb, D. K., Steidel, C. C., Shapley, A. E., Pettini, M., Reddy, N. A., \& Adelberger, K. L. 2006, ApJ, 647, 128
\bibitem[Fazio et al.\ (2004)]{faz04} Fazio, G. G. et al.\ 2004, ApJS, 154, 10
\bibitem[Finkelstein et al.\ (2007)]{fink07} Finkelstein, S. L., Rhoads, J. E., Malhotra, S., Pirzkal, N., \& Wang, J. 2007, ApJ, 660, 1023
\bibitem[F{\"o}rster Schreiber et al.\ (2009)]{for09} F{\"o}rster Schreiber, N. M., et al.\ 2009, ApJ, 706, 1364
\bibitem[Gouliermis et al.\ (2005)]{gou05} Gouliermis, D., Brandner, W., \& Henning, Th. 2005, ApJ, 623, 846
\bibitem[Griffin et al.\ (2010)]{gri10} Griffin, M. J., et al.\ 2010, A\&A, 518, L3
\bibitem[Houck et al.\ (2004)]{hou04} Houck, J. R., et al.\ 2004, ApJS, 154, 18
\bibitem[Hoversten \& Glazebrook (2008)]{hg08} Hoversten, E. A. \& Glazebrook, K. 2008, ApJ, 675, 163
\bibitem[Hoversten \& Glazebrook (2010)]{hg10} Hoversten, E. A. \& Glazebrook, K. 2010, ASP Conference Series, ``UP: Have Observations Revealed a Variable Upper End to the Stellar Initial Mass Function?'', arXiv:1011:3818
\bibitem[Kennicutt (1998)]{kenn} Kennicutt, R. C. 1998, ARAA, 36, 189
\bibitem[Kennicutt et al.\ (2003)]{ken03} Kennicutt, R. C., et al.\ 2003, PASP, 115, 928
\bibitem[Kneib et al.\ (2004)]{kne} Kneib, J.-P., van der Werf, P. P., Kraiberg Knudsen, K., Smail, I., Blain, A., Frayer, D., Barnard, V., \& Ivison, R. 2004, MNRAS, 349, 1211
\bibitem[Kneib et al.\ (2005)]{k05} Kneib, J.-P., Neri, R., Smail, I., Blain, A., Sheth, K., van der Werf, P. P., \& Kraiberg Knudsen, K. 2005, A\&A, 434, 819
\bibitem[Kudritzki et al.\ (2000)]{kud00} Kudritzki, R.-P., et al.\ 2000, ApJ, 536, 19
\bibitem[Lacey et al.\ (2008)]{lac08} Lacey, C. G., Baugh, C. M., Frenk, C. S., Silva, L., Granato, G. L., \& Bressan, A. 2008, MNRAS, 385, 1155
\bibitem[Lee et al.\ (2004)]{lee04} Lee, H.-C., Gibson, B. K., Flynn, C., Kawata, D., \& Beasley, M. A. 2004, MNRAS, 353, 113
\bibitem[Leitherer et al.\ (1999)]{lei99} Leitherer, C. et al.\ 1999, ApJS, 123, 3
\bibitem[Lowenstein \& Mushotzky (1996)]{lm96} Lowenstein, M. \& Mushotzky R. F. 1996, ApJ, 466, 695
\bibitem[Magdis et al.\ (2010)]{mag10} Magdis, G. E., et al.\ 2010, MNRAS, 409, 22
\bibitem[Malhotra \& Rhoads (2002)]{mr02} Malhotra, S. \& Rhoads, J. 2002, ApJ, 565, 71
\bibitem[Muzzin et al.\ (2010)]{muz10} Muzzin, A., van Dokkum, P., Kriek, M., Labb\'e, Cury, I., Marchesini, D., \& Franx, M. 2010, ApJ, 725, 742
\bibitem[Nguyen et al.\ (2010)]{ngu10} Nguyen, H. T., et al.\ 2010, A\&A, 518, L5
\bibitem[Oliver et al.\ (2010)]{oli10} Oliver, S. J., et al.\ 2010, A\&A, 518, L21
\bibitem[Osterbrock (1989)]{ost89} Osterbrock, D. E. 1989, Astrophysics of Gaseous Nebula and Active Galactic Nuclei (Mill Valley, CA: Univ. Science Books) 
\bibitem[Papovich et al.\ (2009)]{pap09} Papovich, C., et al.\ 2009, ApJ, 704, 1506 (P09)
\bibitem[Papovich et al.\ (2011)]{pap11} Papovich, C., Finkelstein, S. L., Ferguson, H. C., Lotz, J. M., \& Giavalisco, M. 2011, MNRAS, 412, 1123
\bibitem[Peng et al.\ (2010)]{pen10} Peng, C. Y., Ho, L. C., Impey, C. D., \& Rix, H.-W. 2010, ApJ, 139, 2097
\bibitem[Pettini \& Pagel (2004)]{pet04} Pettini, M. \& Pagel, B. E. J. 2004, MNRAS, 348, 59
\bibitem[Pilbratt et al.\ (2010)]{pil10} Pilbratt, G. L., et al.\ 2010, A\&A, 518, L1
\bibitem[Popper (1980)]{pop80} Popper, D. M. 1980, ARA\&A, 18, 115
\bibitem[Renzini et al.\ (1993)]{ren93} Renzini, A., Ciotti, L., D'Ercole, A., \& Pellegrini, S. 1993, ApJ, 419, 52
\bibitem[Richard et al.\ (2011)]{ric11} Richard, J., Jones, T., Ellis, R., Stark, D. P., Livermore, R., \& Swinbank, M. 2011, MNRAS, 413, 643
\bibitem[Rieke et al.\ (2004)]{rie04} Rieke, G. H., et al.\ 2004, ApJS, 154, 25
\bibitem[Rieke et al.\ (2009)]{rie09} Rieke, G. H., Alonso-Herrero, A., Weiner, B. J., P\'erez-Gonz\'alez, P. G., Blaylock, M., Donley, J. L., \& Marcillac, D. 2009, ApJ, 692, 556
\bibitem[Rigby et al.\ (2008)]{rig08} Rigby, J. R., et al.\ 2008, ApJ, 675, 262
\bibitem[Salpeter (1955)]{sal} Salpeter, E. E. 1955, ApJ, 121, 161
\bibitem[Sternberg et al.\ (2003)]{ste03} Sternberg, A., Hoffmann, T. L., \& Pauldrach, A. W. A. 2003, ApJ, 599, 1333
\bibitem[Vlahakis et al.\ (2005)]{vla05} Vlahakis, C., Dunne, L., \& Eales, S. 2005, MNRAS, 364, 1253
\bibitem[Willmer et al.\ (2009)]{wil09} Willmer, C. N. A., Rieke, G. H., Le Floc'h, E., Hinz, J. L., Engelbracht, C. W., Marcillac, D., \& Gordon, K. D. 2009, AJ, 138, 146
\bibitem[Young et al.\ (1989)]{you89} Young, J. S., Xie, S., Kenney, J. D. P., \& Rice. W. L.\ 1989, ApJS, 98, 219
\end{thebibliography}
\end{document}